\documentclass[prc,twocolumn,showpacs,floatfix,nofootinbib,preprintnumbers,superscriptaddress,amsmath,amssymb]{revtex4-1}

\usepackage{graphicx}
\usepackage{epsfig}
\usepackage{amsmath,amssymb}
\usepackage{bm}
\usepackage{color}
\usepackage{float}
\usepackage{dcolumn}

\graphicspath{ {images/} }

\begin{document}


\title{Excitation energy dependence of fission in the mercury region}

\author{J. D. McDonnell}

\affiliation{Physics Division, Lawrence Livermore National Laboratory, Livermore, California 94551, USA}
\affiliation{Department of Physics \&
Astronomy, University of Tennessee, Knoxville, Tennessee 37996, USA}

\author{W. Nazarewicz}
\affiliation{Department of Physics \&
Astronomy, University of Tennessee, Knoxville, Tennessee 37996, USA}
\affiliation{Physics Division, Oak Ridge National Laboratory, Oak Ridge, Tennessee 37831, USA}
\affiliation{Institute of Theoretical Physics, University of Warsaw, ul. Ho\.za
69, 00-681 Warsaw, Poland}

\author{J. A. Sheikh}
\affiliation{Department of Physics \&
Astronomy, University of Tennessee, Knoxville, Tennessee 37996, USA}
\affiliation{Physics Division, Oak Ridge National Laboratory, Oak Ridge, Tennessee 37831, USA}
\affiliation{Department of Physics, University of Kashmir, Srinagar, 190 006, India}

\author{A. Staszczak}
\affiliation{Department of Theoretical Physics, Maria Curie-Sk{\l}odowska University, pl. M. Curie-Sk{\l}odowskiej 1, 20-031 Lublin, Poland}
\affiliation{Department of Physics \&
Astronomy, University of Tennessee, Knoxville, Tennessee 37996, USA}

\author{M. Warda}
\affiliation{Department of Theoretical Physics, Maria Curie-Sk{\l}odowska University, pl. M. Curie-Sk{\l}odowskiej 1, 20-031 Lublin, Poland}

\date{\today}

\begin{abstract} 
\begin{description}
 \item[Background]
Recent experiments on $\beta$-delayed fission reported an asymmetric mass yield in  the neutron-deficient nucleus $^{180}$Hg. Earlier experiments in the  mass region $A=190-200$ close to the $\beta$-stability line,  using the $(p,f)$ and $(\alpha,f)$ reactions, observed a more symmetric distribution of fission fragments. While the $\beta$-delayed fission of $^{180}$Hg can be associated with relatively low excitation energy, this is not the case for light-ion reactions, which result in warm compound nuclei.
The low-energy fission of $^{180,198}$Hg has been successfully described by theory in terms of strong shell effects in pre-scission configurations
associated with di-nuclear structures.  
 
\item[Purpose]
To elucidate the roles of proton and neutron numbers and excitation energy in determining symmetric and asymmetric fission yields, we compute and analyze the isentropic potential energy surfaces of $^{174,180,198}$Hg and $^{196, 210}$Po.
\item[Methods]
We use the finite-temperature superfluid nuclear density functional theory, for excitation energies up to $E^* = 30$\,MeV and zero angular momentum. For our theoretical framework, we consider the Skyrme energy density functional SkM$^*$ and a density-dependent pairing interaction.
\item[Results]
For $^{174,180}$Hg, we predict fission pathways consistent with asymmetric fission at low excitation energies, with the symmetric fission pathway opening very gradually  as excitation energy is increased.  For $^{198}$Hg and $^{196}$Po, we expect the nearly-symmetric fission channel to dominate.  $^{210}$Po shows a preference for a slightly asymmetric pathway at low energies, and a preference for a symmetric pathway at high energies.  
\item[Conclusions]
Our self-consistent theory suggests that excitation energy weakly affects the fission pattern  of the nuclei considered. The transition from the asymmetric fission in the proton-rich nuclei to a more symmetric fission in the heavier isotopes is governed by the shell structure of pre-scission configurations.
\end{description}

\end{abstract}

\maketitle

{\it Introduction} --- The recent experimental discovery of asymmetric fission in $^{180}$Hg via the $\beta$-decay of $^{180}$Tl initially came as a surprise \cite{AndreyevPRL105,andreyev_colloquium:_2013,Elseviers2013}. The mass distribution in  low-energy fission of actinides has been explained in terms of shell effects in nascent fragments \cite{BrackRevModPhys44,Wilkins76,Pashkevich88,MollerNature409,Staszczak09}. In particular, the doubly-magic nuclei $^{132}$Sn and $^{78}$Ni are expected to play a key role in explaining the observed distributions \cite{Gorodisskiy2002,Schmidt08}.
Taking this point literally, the most likely division of $^{180}$Hg would seem to be two symmetric fragments resembling semi-magic $^{90}$Zr.  However, the experiment observed \cite{AndreyevPRL105} a more likely split of $^{100}$Ru and $^{80}$Kr, neither of which is near magicity or particularly strongly stabilized by shell effects.

A theoretical description of this puzzle has been offered in terms of both
macroscopic-microscopic  \citep{AndreyevPRL105, MollerPRC85,IchikawaPRC86,andreev_mass_2012} and self-consistent   \cite{WardaPRC86} approaches. The conclusion that has emerged is  
that the main factor determining  the mass split in fission are shell effects at pre-scission configurations, i.e., between saddle
and scission~\cite{Schmidt08,MulginNPA1998} (see also Ref.~\cite{PanebiancoPRC86} for a scission-point description).

It is important to realize that the recent and previous experiments that study fission in the mercury-lead region populate the fissioning nucleus  at a non-zero excitation energy -- the electron capture considered in Ref.~\cite{AndreyevPRL105} produced $^{180}$Hg with up to $E^*=10.44$\,MeV of excitation energy (limited by the electron-capture $Q$-value  of the precursor $^{180}$Tl) while even  higher excitation energy ranges  were explored in the earlier 
$(p,f)$ and $(\alpha,f)$ studies \cite{MulginNPA1998,itkis1990mass,*itkis1991low}.
The Brownian shape motion model of Ref.~\cite{MollerPRC85}, based on zero-temperature macroscopic-microscopic potential energy surfaces, explored the effects of the imparted excitation energy on the fission yields and has been quite successful in explaining existing experimental data. However, a self-consistent study considering the thermal effects on the fission pathway in this mass region is lacking. To fill this gap, in this work we extend the self-consistent study of Ref.~\cite{WardaPRC86} to explore the effects of excitation energy on the expected mass yields. We employ the finite-temperature density functional theory (FT-DFT) to study the  evolution of the fission pathway with increasing $E^*$.  
Following Ref.~\cite{WardaPRC86}, we study the potential energy surfaces (PESs) of $^{180}$Hg and $^{198}$Hg to get a sense of the trend with neutron number.  We also present fission pathways for $^{196, 210}$Po, which are of recent experimental interest. Furthermore, we investigate the fission of  $^{174}$Hg, which was predicted in Ref.~\cite{andreev_isospin_2013} to favor symmetric fission and 
in Ref.~\cite{MollerPRC85} to gradually change from symmetric at low energies to asymmetric at higher energies --- the opposite of the familiar trend in the actinides.

{\it The Model} ---  To study the PESs as a function of $E^*$, we employ the superfluid 
mean-field theory \cite{GoodmanNPA352,EgidoNPA451} implemented for the Skyrme DFT in Refs.~\cite{PeiPRL102,SheikhPRC80,McDonnellFissionProc2009,mcdonnell_third_2013}. To solve the  finite-temperature  DFT equations, we employ the 
symmetry-unrestricted Skyrme DFT solver  HFODD \cite{SchunckCPC2012}.  
We use a basis that employs the lowest 1140 stretched deformed harmonic oscillator states originating from 31 major shells.  Our previous studies (see, e.g., Ref.~\cite{mcdonnell_third_2013} and references therein) indicate that this basis size represents a good compromise between accuracy and computation time.  

We constrain the quadrupole moment $Q_{20}$ and the octupole moment $Q_{30}$ with the augmented Lagrangian method to obtain the PES \cite{staszczak_augmented_2010}.  To obtain smooth PESs, we use cubic spline interpolation.  To construct one-dimensional least-energy pathways, we have taken two approaches.  
The first approach was to initially constrain $Q_{20}$ along the pathway as well as $Q_{30}$ at some non-zero number (a value of $10$\,b$^{3/2}$ is typically sufficient).  We would then release the $Q_{30}$ constraint, allowing the pathway to fall into the least-energy trajectory.  
The second approach was to directly scan the two-dimensional PESs.  For each value of $Q_{20}$, we would scan in the $Q_{30}$ direction for the least-energy point.  The locus of these points formed the least-energy pathway.  In addition, we sought alternative fission pathways (either more or less symmetry-breaking) by restricting the scan to a subset of $Q_{30}$ values.  
Both approaches led to least-energy pathways in excellent agreement with each other.  

The finite-temperature DFT equations are obtained from the minimization of the grand canonical potential, so that the free energy $F = E - TS$ is formally calculated at a fixed temperature $T$. Since the system is not in contact with a heat bath, the fission process is not isothermal. However,  since the large-amplitude collective motion during fission  is slower than the single-particle motion, it is reasonable to treat fission as an adiabatic isentropic  process \cite{PeiPRL102,NazarewiczNPA1993}.   We calculate the free energy for a fixed temperature as a function of the collective coordinates, understanding that relative quantities such as barrier heights  identically match those obtained from a calculation of internal energy at fixed entropy  \cite{Diebel1981,Faber1984,PeiPRL102}. This Maxwell relation has  been verified numerically in the self-consistent calculations of Ref. \cite{PeiPRL102}.

We map the excitation energy of the nucleus $E^*$ to the fixed temperature $T$ via
\begin{equation}
E^* (T) = E_{\mathrm{g.s.}}(T) - E_{\mathrm{g.s.}}(T = 0),
\end{equation}
where $E_{\mathrm{g.s.}}(T)$ is the minimum energy of the nucleus at temperature $T$.  This corresponds well to the excitation energy of a compound nucleus \cite{PeiPRL102,SheikhPRC80}. To study shell effects in pre-scission configurations,   we calculate the shell correction energies $\delta E^{\mathrm{sh}}$ at $T=0$ according to the procedure described in
Refs.~\cite{VertsePRC61,NikolovPRC83} with the smoothing width parameters
$\gamma_p = 1.66, \gamma_n = 1.54$ (in units of $\hbar\omega_0 = 41/A^{1/3}$ MeV) and the curvature correction $p = 10$.

The nuclear interaction in the particle-hole channel has been approximated through the SkM$^*$ parametrization \cite{BartelNPA386} of the Skyrme energy density functional.  This traditional  functional achieves realistic surface properties in the actinides, allowing a good description of the evolution of the energy with deformation \cite{Staszczak09,WardaPRC86}.  
In the particle-particle channel, we use the density-dependent mixed-pairing interaction \cite{Dobaczewski02}. All calculations were performed
with a quasiparticle cutoff energy of $E_{\rm cut} = 60$\,MeV. The pairing strengths
$V_{\tau 0}$ ($\tau=n, p$) are chosen to fit the pairing gaps determined from experimental odd-even mass differences in $^{180}$Hg \cite{AME2003}.  For SkM$^*$ EDF, the pairing strengths are $V_{n0} = -268.9$\,MeV and
$V_{p0} = -332.5$\,MeV.

In this work, we have chosen to focus our attention on the effect that internal excitation energy has on mass yield.  We do not consider the sharing of projectile energy between nuclear excitation and nuclear rotation.  While the earlier experiments with projectiles would involve a great deal of angular momentum imparted to the fissioning nucleus, the more recent experiments with $\beta$-delayed fission achieve a low angular momentum for the fissioning nucleus.  

\begin{figure}[tbh]
  \includegraphics[width=1.0\columnwidth]{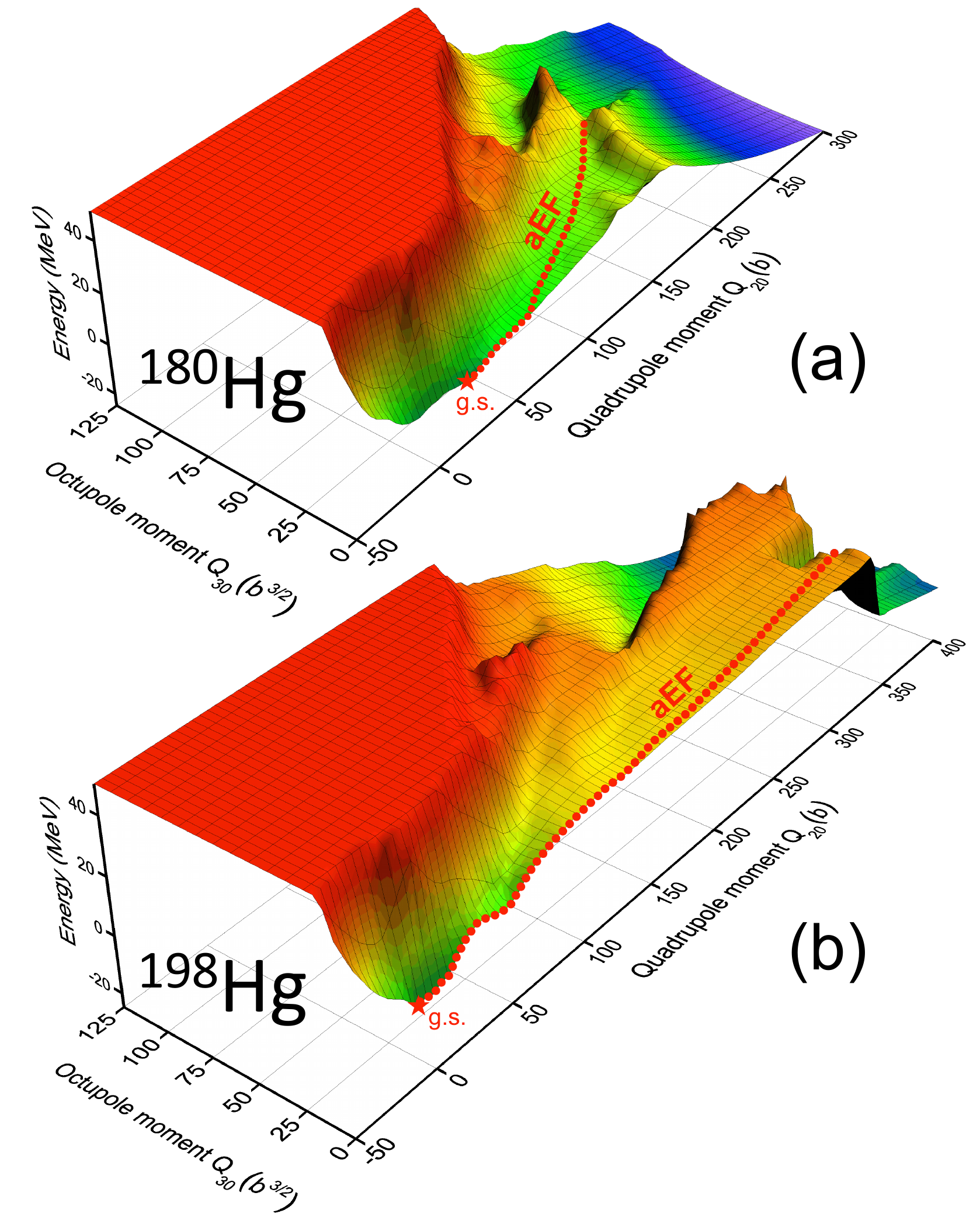}
  \caption[T]{\label{fig:hg3D}
  (Color online) Ground-state potential-energy surfaces for (a) $^{180}$Hg and (b) $^{198}$Hg in  the $(Q_{20}, Q_{30})$ plane calculated in HFB-SkM$^*$. The static fission pathway aEF corresponding to asymmetric elongated fragments
is marked.}
 \end{figure}

\begin{figure}[tbh]
  \includegraphics[width=0.8\columnwidth]{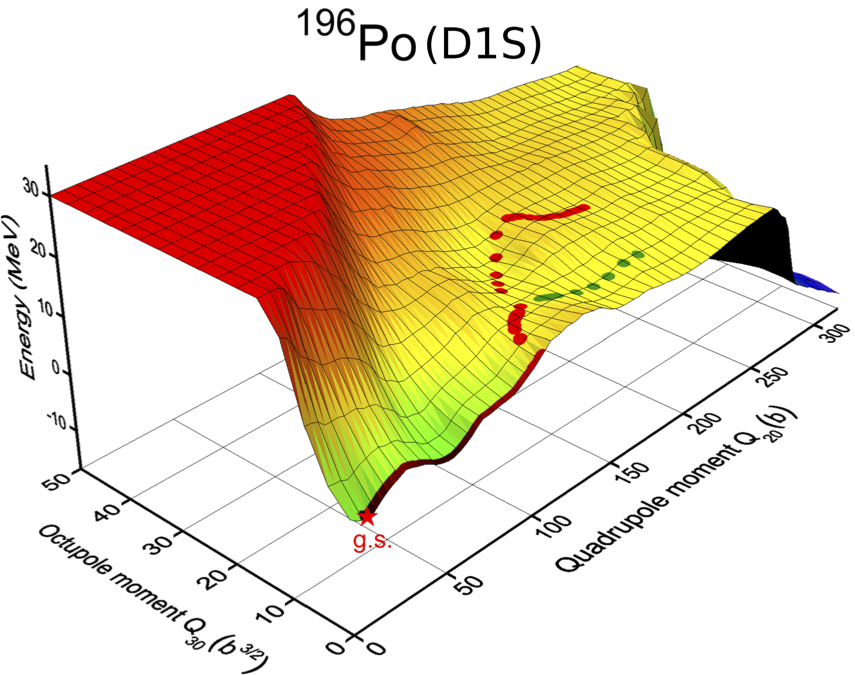}
  \caption[T]{\label{fig:po3D}
   (Color online) Similar to Fig.~\ref{fig:hg3D}, but for $^{196}$Po in HFB-D1S. Two competing fission pathways corresponding to  different mass asymmetry are marked.}
 \end{figure}

{\it Results} ---  To recall the global features of the PESs predicted with HFB-SkM$^*$ in the Hg region \cite{WardaPRC86}, in Fig.~\ref{fig:hg3D} we show
the results for $^{180}$Hg and $^{198}$Hg at zero-temperature.  This exploration of a very large configuration space illustrates the static fission paths available to each nuclide.  In both cases, the reflection-asymmetric path
 corresponding to elongated fission
fragments (aEF) branches
away from the symmetric valley to ultimately
pass through the mass-asymmetric scission point.
For $^{180}$Hg, a steep ridge separating the 
path aEF from the fusion valley at $Q_{20} \approx 175$\,b can be seen.  The $^{100}$Ru$/ ^{80}$Kr split for $^{180}$Hg corresponds to $Q_{30} \approx 30$\,b$^{3/2}$ at $Q_{20} \approx 300$\,b.  As was discussed in greater detail in Ref.~\cite{WardaPRC86}, the PES of $^{180}$Hg exhibits an asymmetric fission pathway that is clearly separated from the symmetric fission pathway by a barrier.  
In contrast, the PES of $^{198}$Hg is rather flat in the pre-scission region:   symmetric and asymmetric  pathways have almost equal energy. Therefore,
the HFB-SkM$^*$  model explains the transition from asymmetric fission in $^{180}$Hg toward a more symmetric
distribution of fission fragments in $^{198}$Hg that has been seen experimentally.
%
%
Stimulated by recent experimental work \cite{Ghys,Liberati,Poexps,*Poexps1}, we have also examined the PES of $^{196}$Po by extending the calculations of Ref.~\cite{WardaPRC86} with the D1S interaction \cite{BergerNPA428}.  Figure~\ref{fig:po3D} shows that the least-energy fission pathway   is nearly reflection-symmetric ($Q_{30} \approx 10$\,b$^{3/2}$).  Our calculations also predict a  secondary, more asymmetric pathway.
Beyond $Q_{20} = 250$\,b, the PES appears to flatten so that mildly asymmetric fission pathways compete with symmetric pathways in that region.  
 
\begin{figure}[tbh]
  \includegraphics[width=0.8\columnwidth]{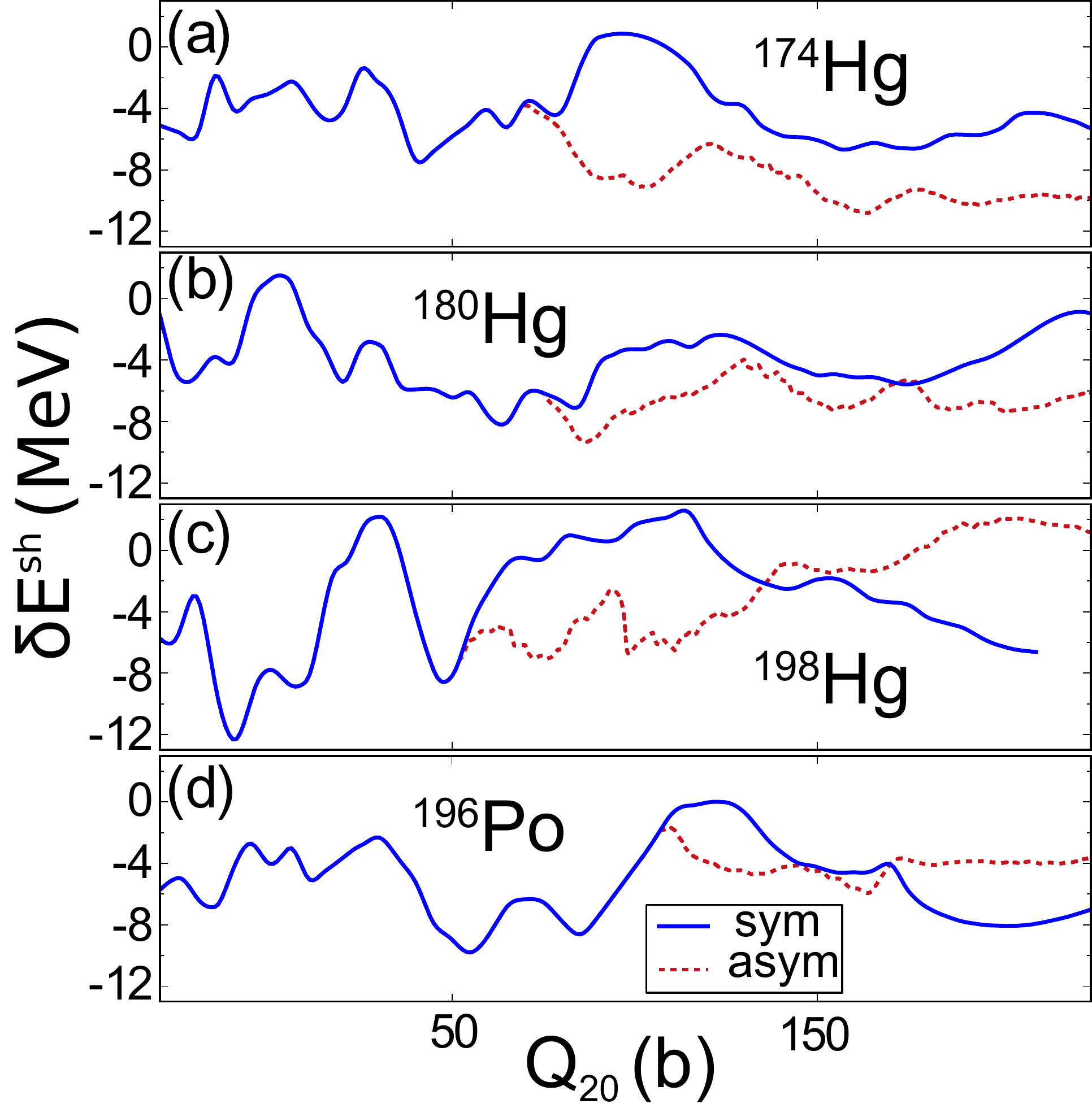}
  \caption[T]{\label{fig:hgshcontours}
  (Color online) The total shell correction energy at $T=0$ in  $^{174,180,198}$Hg and $^{196}$Po along the symmetric (solid line, $Q_{30}=0$) and asymmetric (dashed line)  fission pathways.}
 \end{figure}

Figure~\ref{fig:hgshcontours} shows the total shell correction energy $\delta E^{sh}$ along the symmetric ($Q_{30}=0$) and asymmetric fission paths in   $^{174,180,198}$Hg and $^{196}$Po, each calculated with SkM$^*$. This figure nicely corroborates the results in Fig.~\ref{fig:hg3D}: the preference for the asymmetric pathway in $^{180}$Hg is driven by shell effects in pre-scission configurations. In addition, one can see that shell effects drive $^{174}$Hg towards asymmetric splits, as well as the competition from more mass-symmetric pathways in $^{198}$Hg and  $^{196}$Po.  

  \begin{figure}
  \includegraphics[width=0.8\columnwidth]{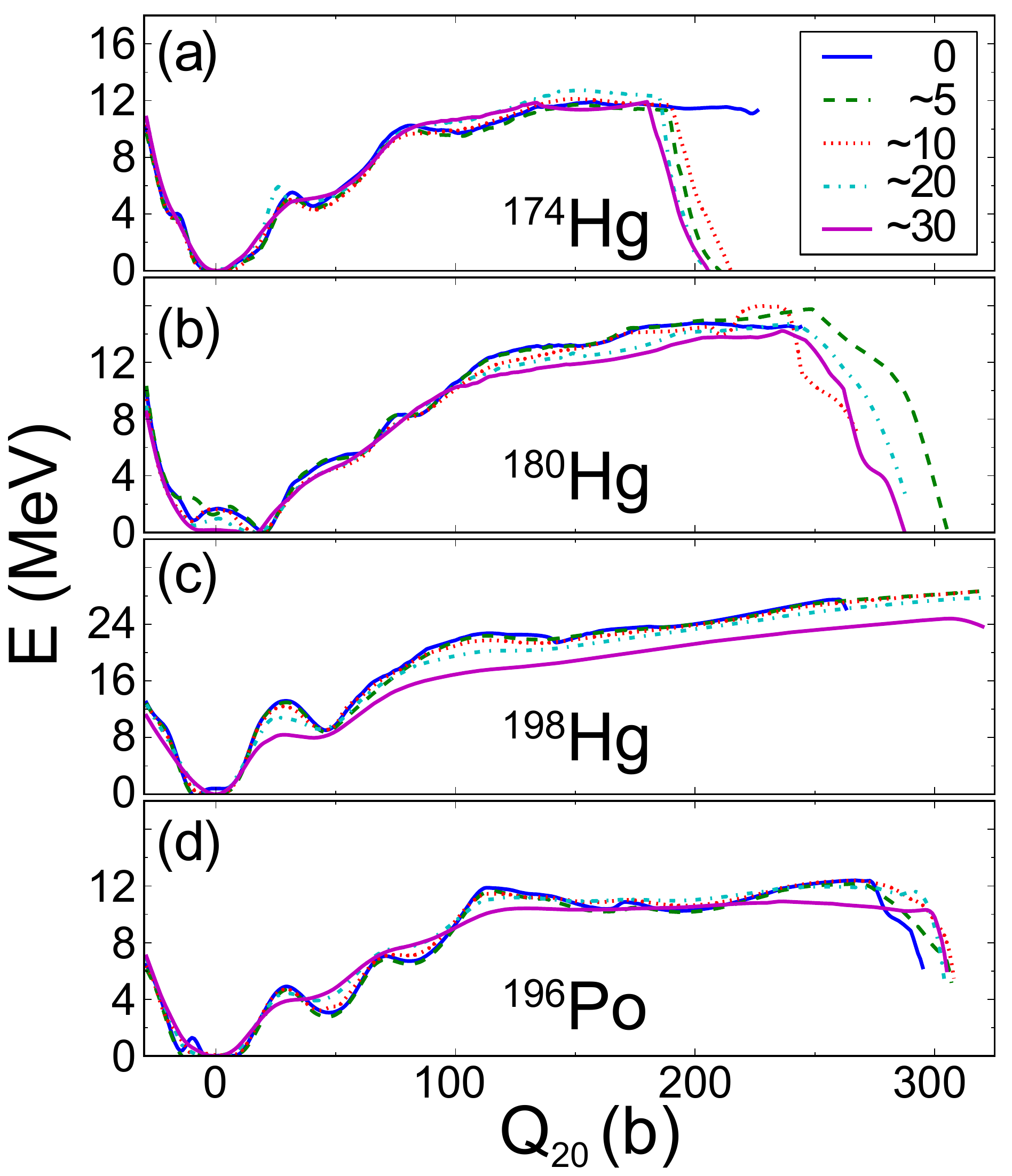}
  \caption[T]{\label{fig:hg1dplots}
  (Color online) Potential-energy curves for $^{174,180,198}$Hg and $^{196}$Po at different values of excitation energy (in MeV).}
 \end{figure}
In the following, we now turn to study the evolution of the symmetry of the fission yield for these mercury and polonium isotopes with increasing excitation energy.  
First, we wish to verify that our finite-temperature theory successfully reproduces (i) asymmetric fission around $^{180}$Hg at relatively low excitation energies,  and (ii) nearly-symmetric fission around $^{198}$Hg as observed in the $(p,f)$ and $(\alpha,f)$ studies \cite{MulginNPA1998,itkis1990mass,*itkis1991low}.  In Fig.~\ref{fig:hg1dplots}, we show  the predicted evolution of the  potential-energy curves along the static fission pathways of $^{174,180,198}$Hg and $^{196}$Po at different  excitation energies chosen across the range of energies explored  experimentally. 
We also performed two-dimensional calculations  in  the $(Q_{20}, Q_{30})$ plane to trace the evolution of the mass asymmetry with temperature, as well as to ascertain that we are following a continuous static fission pathway. 
  \begin{figure}
  \includegraphics[width=1.0\columnwidth]{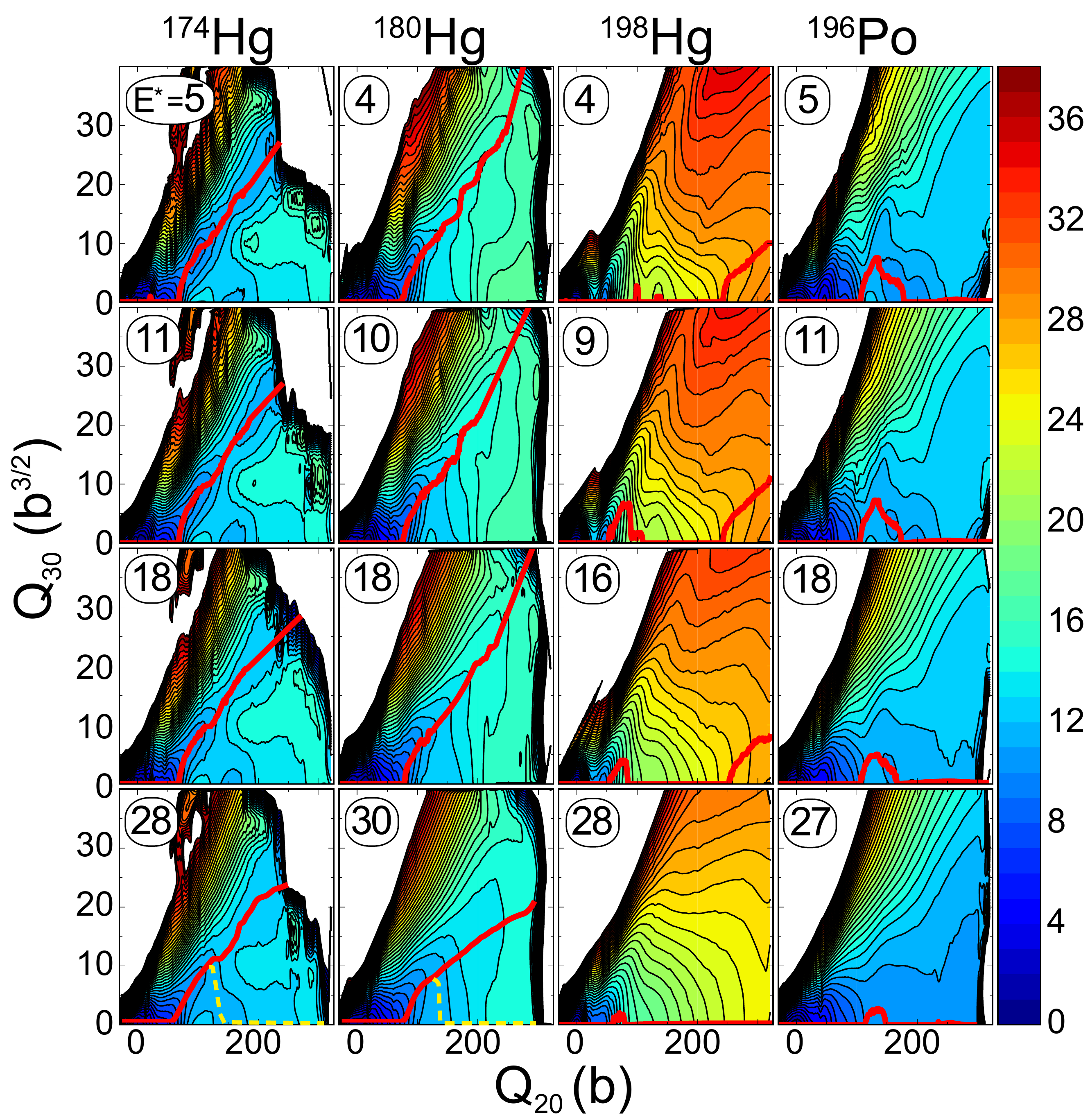}
  \caption[T]{\label{fig:hgFTcontours}
  (Color online) Potential energy surfaces for $^{174,180,198}$Hg and $^{196}$Po
  in  the $(Q_{20}, Q_{30})$ plane at different values of $E^*$ indicated (in MeV).  The contour lines are separated by $1$\,MeV.  The lowest static  (least-energy) fission pathway is marked with a thick red line. The secondary fission pathways in $^{174,180}$Hg are marked with a yellow dashed line at the highest excitation energy.}
\end{figure}

The lighter nuclei $^{174}$Hg and $^{180}$Hg follow the same trends: asymmetric fission at low energies, with a decreasing barrier for symmetric fission as excitation energy increases. In the case of $^{174,180}$Hg, a secondary, symmetric pathway branches out slightly beyond $Q_{20}=100$\,b (indicated by a yellow dashed
line in Fig.~\ref{fig:hgFTcontours} at the highest excitation energy).
At $E^*\approx 30$\,MeV,  these symmetric and asymmetric pathways are quite close in energy.  We note that the alternative, symmetric pathway exists at lower excitation energies, but that it presents a barrier about $1-2$\,MeV higher than the asymmetric pathway.  

For $^{198}$Hg and $^{196}$Po, we see in Fig~\ref{fig:hgshcontours} that shell effects drive each system towards symmetric fission at zero temperature.  The results in Fig.~\ref{fig:hgFTcontours} affirm the preference for fission pathways that only mildly break reflection symmetry.  The static  fission pathway for $^{198}$Hg corresponds to far smaller values of $Q_{30}$ than those for $^{174,180}$Hg.  For $^{196}$Po, at each excitation energy the PES is rather flat in the $Q_{30}$ direction and appears to permit some symmetry-breaking pathways, but the symmetric pathway is consistently lowest in energy.

Whether the symmetric or asymmetric pathway is favored at each excitation energy would be determined decisively with a dynamic, finite-temperature calculation of the system's path of least action, which requires a determination of a finite-temperature collective inertia. Based on our static calculations, however, we predict that  the symmetric yield should gradually increase with increasing $E^*$. In $^{174}$Hg and $^{180}$Hg, we expect asymmetric fission to dominate at all excitation energies considered. 
Our prediction for $^{174}$Hg is at variance with that of the macroscopic-microscopic model \cite{MollerPRC85}; they predict that the yield distribution in this nucleus  should become {\it more} symmetric with {\it decreasing} $E^*$, suggesting a preference for a nearly-symmetric fission pathway in their model.  Similarly, in Ref.~\cite{andreev_isospin_2013} the authors use the scission point model to predict that the $^{174}$Hg yield is symmetric at each excitation energy.  

For $^{198}$Hg, we expect a fairly symmetric pattern of fission yields. Finally, for $^{196}$Po we predict a dominance of symmetric fission with some competition from a secondary asymmetric pathway, as seen in Figs.~\ref{fig:po3D} and \ref{fig:hgFTcontours}. 

\begin{figure}[tbh]
  \includegraphics[width=1.0\columnwidth]{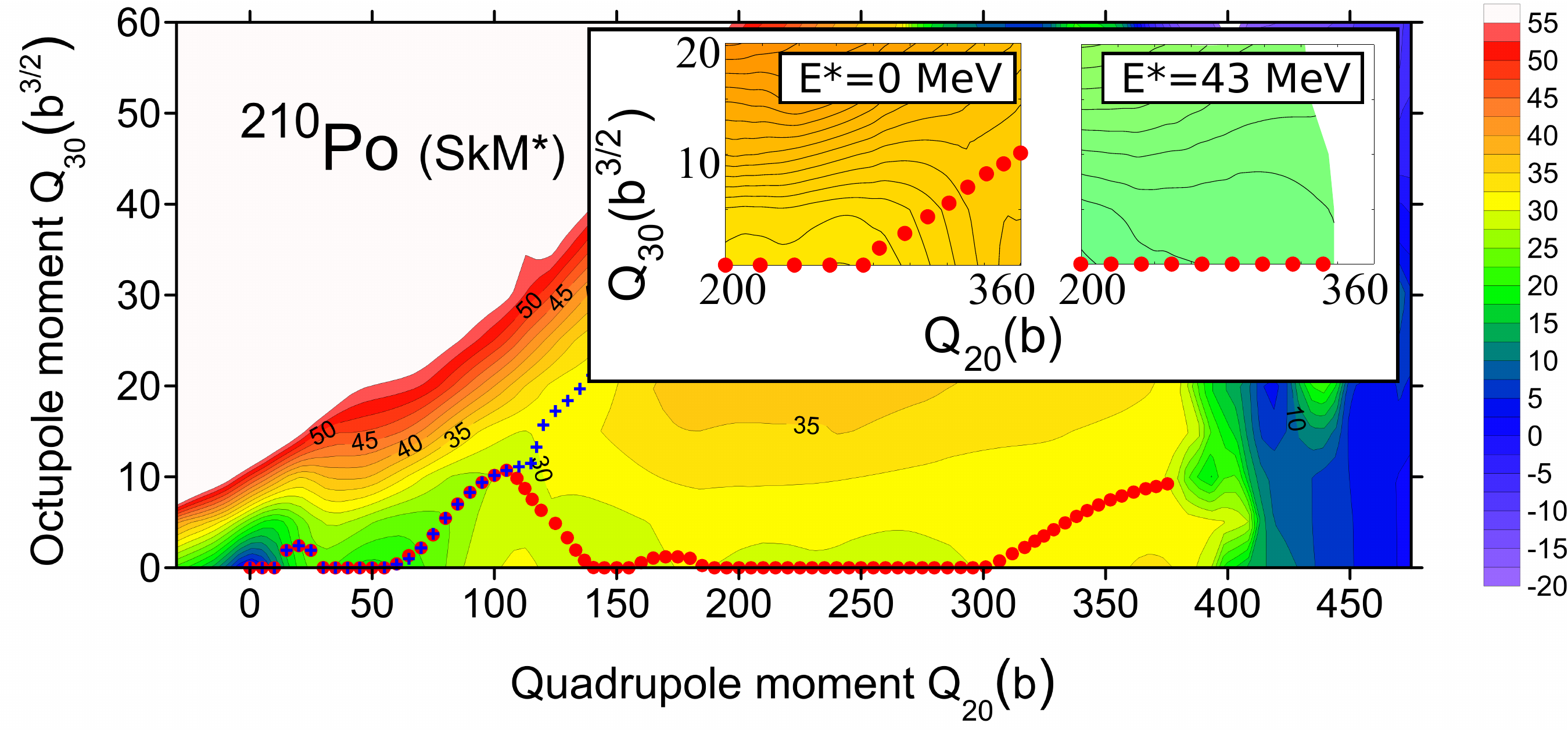}
  \caption[T]{\label{fig:po210-3D}
   (Color online) Potential energy surface for $^{210}$Po in HFB-SkM$^*$. Both the dominant, mildly symmetry-breaking pathway and a secondary, more severely symmetry-breaking pathway are marked.  The inset offers a closer look at a pre-scission region $(200 < Q_{20} < 360, 0 < Q_{30} < 20)$ for $E^* = 0$\,MeV and for a higher excitation energy $E^*=43$\,MeV. }
 \end{figure}

As a final case, we present predictions for $^{210}$Po.  This nucleus was studied in Ref.~\cite{MulginNPA1998}, and symmetric fission has been observed to dominate, especially at higher excitation energies.  To see whether our model reproduces this feature, in Fig.~\ref{fig:po210-3D} we show the potential energy surface for $^{210}$Po in the $(Q_{20}, Q_{30})$ plane. At $E^* = 0$\,MeV, the dominant fission pathway favors a slight mass asymmetry, diverging from a perfectly symmetric pathway at $Q_{20}\approx 290$\,b.  With increasing excitation energy,  however, a transition is observed towards the symmetric fission pathway. As seen in the inset,  at $E^*=43$\,MeV, only the symmetric pathway remains; this is consistent with the temperature damping of shell effects discussed in Ref.~\cite{MulginNPA1998}.   At $E^*=0$\,MeV, we also predict a more severely symmetry-breaking fission pathway (cluster radioactivity), which lies at higher energy and diverges from the dominant pathway at $Q_{20}\approx 100$\,b. 

{\it Conclusions} ---  Nuclear fission in the mercury region has been studied with the superfluid FT-DFT, exploring the evolution of 
fission pathways with increasing excitation energy.  This is a necessary ingredient for a theoretical description of the recent experimental discoveries
in this region.
 
Our potential-energy surfaces show the proclivity of both $^{174}$Hg and $^{180}$Hg towards an asymmetric fission pathway when the nucleus has low excitation energy.  As excitation energy is increased, we see a gradual lowering of the barrier to the symmetric fission valley.  Because of relatively large macroscopic barriers ($\approx 12-24$\,MeV), this trend with excitation energy is very gentle, so that we predict asymmetric fission to dominate in both isotopes at least up to $30$\,MeV.  This is unlike the actinides, where macroscopic barriers are much smaller, and the calculations demonstrate a stronger dependence on excitation energy \cite{SheikhPRC80}, in accord with experiment.

For $^{174}$Hg, our prediction differs from that of Ref.~\cite{MollerPRC85} and Ref.~\cite{andreev_isospin_2013} -- we predict asymmetric fission at each excitation energy, rather than symmetric fission at low energies.  It would be very interesting for future experiments to shed more light on this issue.

For $^{174,180}$Hg in particular, the potential energy surface is soft in the $Q_{30}$ coordinate for a large range of excitation energy.  It would be very interesting for future work to study how increased excitation energy affects the collective motion of the system --- such a dynamical study would aid in pinpointing a theoretical prediction for the energy where the system transitions from asymmetric to symmetric fission.

The techniques of the present study can be  extended to explore the findings of continuing experimental investigations in the neutron-deficient Hg-Pb region \cite{andreyev_-delayed_2013,lane_-delayed_2013}.

The theoretical picture of fission in the mercury-polonium region that has been emerging since the experimental discovery of the asymmetric fission mode in $^{180}$Hg demonstrates subtleties not immediately obvious from past fission studies in the actinides.  A thorough examination of potential-energy surfaces with many degrees of freedom was needed to reliably predict the most likely fission path for the mercury and polonium isotopes.  
In particular, the transition with neutron number from asymmetric fission in $^{180}$Hg to symmetric fission in $^{198}$Hg illustrates an intricacy of nuclear fission that must be captured by any reliable theory.  
That the FT-DFT fission model, whose only input is the nuclear energy density functional, captures these salient features is  encouraging.

\begin{acknowledgments}
Useful discussions with A. Andreyev and N. Schunck 
are gratefully acknowledged. 
This work was finalized during the  Program INT-13-3 ``Quantitative Large Amplitude Shape Dynamics: 
fission and heavy ion fusion" at the National Institute for Nuclear Theory in Seattle; it was
supported by
the U.S. Department of Energy under
Contract No.  DE-FG02-96ER40963
 (University of Tennessee), No. DE-NA0001820 (the Stewardship Science Academic Alliances program), and No.
DE-SC0008499    (NUCLEI SciDAC Collaboration).   This work was partly supported by the Polish National Science Center Grant
No.~DEC-2011/01/B/ST2/03667. J.M. was funded by a NNSA SSGF Fellowship under grant No. DE-FC52-08NA28752.  This work was performed under the auspices of the U.S. Department of Energy by the Lawrence Livermore National Laboratory under Contract No. DE-AC52- 07NA27344. Computational resources were provided through an INCITE award ``Computational
Nuclear Structure" by the National Center for Computational Sciences (NCCS), 
and by the National Institute for Computational Sciences (NICS).

\end{acknowledgments}


\bibliographystyle{apsrev4-1}
\bibliography{jmcdonnell-hg180}

\end{document}